\input epsf.sty

\documentclass[twocolumn,showpacs,preprintnumbers,amsmath,amssymb]{revtex4}
\usepackage{graphicx}

\begin{document}

\title{ARPES Study of the Metal-Insulator Transition in Bismuth Cobaltates}
\author{
        Z. Yusof,$^1$\cite{ZY} B.O. Wells,$^1$
        T. Valla,$^2$ P.D. Johnson,$^2$
      A.V. Fedorov,$^{2}$\cite{AF}
      Q. Li,$^{3}$
        S.M. Loureiro,$^{4}$ and R.J. Cava,$^{4}$
        }
\address{
         (1) Department of Physics, University of Connecticut, 2152 Hillside Road U-46, Storrs, CT 06269-3046.\\
         (2) Physics Department, Bldg. 510B, Brookhaven National Laboratory, Upton, NY 11973-5000.\\
         (3) Division of Materials Sciences, Brookhaven National Laboratory, Upton, NY 11973-5000.\\
         (4) Department of Chemistry and Materials Institute, Princeton University, Princeton, NJ 08544.
         }

\begin{minipage}[t]{7.0in}
\begin{abstract}
We present an angle-resolved photoemission spectroscopy (ARPES)
study of a Mott-Hubbard-type bismuth cobaltate system across a
metal-insulator transition. By varying the amount of Pb
substitution, and by doping with Sr or Ba cation, a range of
insulating to metallic properties is obtained. We observe a
systematic change in the spectral weight of the coherent and
incoherent parts, accompanied by an energy shift of the incoherent
part. The band dispersion also shows the emergence of a weakly
dispersing state at the Fermi energy with increasing conductivity.
These changes correspond with the changes in the
temperature-dependent resistivity behavior. We address the nature
of the coherent-incoherent parts in relation to the peak-dip-hump
feature seen in cuprates superconductors.
\end{abstract}
\end{minipage}

\maketitle
\narrowtext

A central issue in condensed matter physics is the metal-insulator
transition (MIT) in strongly-correlated electron systems, an issue
in such areas as high-$T_{c}$ superconductors and CMR manganites.
An important step to understand the nature of charge transport in
such materials would be the ability to understand and predict the
single-particle excitation spectrum - the spectral weight function
- as measured by angle-resolved photoemission spectroscopy
(ARPES). In most materials, there are a variety of
complications to the ARPES spectra making a comparison to theory
difficult. For example, in high-$T_{c}$ superconductors the most
successful samples have multiple bands crossing the Fermi energy
$E_{F}$, gaps at $E_{F}$, and anisotropic spectral lineshapes that
differ strongly in different areas of the Brillouin zone. It would
be helpful to study a compound with similar many-body physics and
a similar MIT but with a simpler spectral function. Here we report
on the ARPES-derived spectral functions of the layered oxide
Bi$_{2-x}$Pb$_{x}$M$_{2}$Co$_{2}$O$_{8}$, with M = Sr, Ba, Ca. The
spectral functions of these compounds appear to be simple enough
to allow for a qualitative comparison to popular many-body
models and indicate important trends as a function of temperature
and doping.

Doping can cause a MIT in a $d$-band Mott insulator in two ways
\cite{Fujimori}. The first is charge doping to change the
electron filling level. The second is iso-electronic chemical
doping to alter the  $U/W$ ratio, where $U$ is the on-site coulomb
repulsion and $W$ is the bandwidth. Both types of doping can be
done in this family of cobalt oxides. Theoretically, models for
this transition typically fall into one of two categories. The
first starts from a description of the insulating state based upon
the Hubbard Hamiltonian. In the Hubbard picture \cite{Hubbard},
reducing $U/W$ causes the gap between the upper and lower Hubbard
bands to close and a transition to a metal occurs when the bands
overlap. The other category is the Brinkman-Rice picture that
starts from the metallic phase. Here, the quasiparticle (QP) band
at $E_{F}$ narrows and loses spectral weight as the transition is
approached \cite{Brinkman}. An approach that bridges these two
approaches is the Dynamical Mean Field Theory (DMFT)
\cite{Georges}. As we show here, spectral functions for the DMFT
are remarkably similar to those we have measured in these cobalt
oxides.

The structure of the layered oxides of the type
Bi$_{2-x}$Pb$_{x}$M$_{2}$Co$_{2}$O$_{8}$ consist of 
rock salt layers of Bi(Pb)MO$_{2}$ interspersed with misfit
hexagonal layers of CoO$_{2}$
\cite{Yamamoto,Tsukuda,Cava}.  In general,
samples with M = Ba are more conducting than those with M = Sr;
such doping is charge neutral and should be considered as
primarily changing $U/W$. For a given M atom species, increasing
the amount of Pb substituted for Bi increases the conductivity. It
is believed that adding Pb increases the number of carriers, thus
is primarily charge doping \cite{Yamamoto}.

The Co ions appear to be in the low spin state as measured by
susceptibility\cite{Tsukuda} and core level photoemission
\cite{Mizokawa}. The Co oxdation state in a Pb-free M = Sr sample
has been calculated to be 3.33 \cite{Yamamoto2}. Thus it is
unlikely that any of the compounds studied here are at
half-filling, though many are strongly insulating. The more
metallic samples are typically in-plane metallic and out-of-plane
insulating at room temperature, metallic in all directions below
some crossover temperature near 150K, and insulating in all
directions at the lowest temperatures \cite{Tsukuda,Valla-Nature}.
Our previous ARPES work has shown that emergence of a sharp peak
at $E_{F}$ tied to a change in the $c$-axis resistivity from
insulating ($d\rho/dT<0$) to metallic-like ($d\rho/dT>0$)
\cite{Valla-Nature}. For the M = Sr and no Pb doping, the
insulating low temperature phase resistivity shows nearly
activated behavior. In the same family, for Pb content over 0.4,
the low temperature resistivity indicates variable range hopping
\cite{Yamamoto2}.

Single crystals of Bi$_{2-x}$Pb$_{x}$M$_{2}$Co$_{2}$O$_{8}$ with M
= Sr or Ba, were synthesized using the flux technique as described
elsewhere \cite{Loureiro}. For convenience we list the samples as
$M/x$ in order to describe both the cation and the Pb doping
level. Thus $Sr/0.52$ indicates
Bi$_{1.48}$Pb$_{0.52}$Sr$_{2}$Co$_{2}$O$_{8}$. All ARPES
measurements were performed at Beamline U13UB of the 
National Synchrotron Light Source, with photon
energies of 15 eV and 21.2 eV. The end station includes a Scienta
SES-200 hemispherical analyzer equipped for simultaneous
collection of photoelectrons as a function of energy and angle.
The crystals were cleaved \textit{in situ} under vacuum with a
base pressure of $1\times 10^{-10}$ Torr. The total energy
resolution was $\sim 15$ meV and momentum resolution better than
0.02 A$^{-1}$.

\begin{figure}
\includegraphics[clip,width=3.25in]{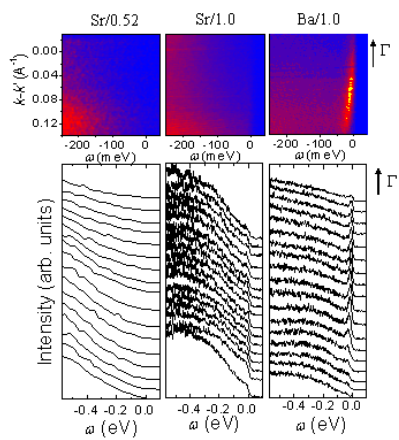}
\caption{(color) ARPES intensity for Sample $Sr/0.52$ ($T=50$K,
left column), $Sr/1.0$ ($T=40$K, middle), and $Ba/1.0$ ($T=30$K,
right). The top panels are the 2D intensity as a function of
energy $\omega$ and momentum $k$ measured from $k'=k_{F}$, where
$k'=k_{F}$, the Fermi wavevector for Sample $Sr/1.0$. The bottom
panels are the corresponding EDCs over the same $k$ range.}
\label{fig1}
\end{figure}

Fig. 1 shows the photoemission intensity near $E_{F}$ along the
rock salt cube-edge direction for a set of cobalt oxide samples.
There is some variation in the dispersion in other directions but
the lineshapes are essentially the same. The data is plotted both
as two dimensional intensity maps as collected as well as energy
dispersion curves (EDC's) at several $k$ values. The left most
panel is sample $Sr/0.52$ which is the most insulating of those
examined here. Not surprisingly, there is no weight at $E_{F}$ and
a very broad feature with no discernible dispersion. The middle
panel is sample $Sr/1.0$. This sample is charge doped with respect
to $Sr/0.52$ shown in the left most panel and is barely metallic,
with details of the conductivity below. This sample has developed
a barely discernible peak at $E_{F}$ with slight dispersion. The
rightmost panel is $Ba/1.0$. This is the most metallic of those
studied here. The change from the $Sr/1.0$ in the middle panel
involves substitution of Ba for Sr, and thus decreasing $U/W$. In
the near $E_{F}$ region we see a more intense peak with slightly
greater dispersion. The existence of a sharp, slowly dispersion
peak in the metallic samples that weakens as the insulator is
approached is reminiscent of the Brinkman-Rice state approaching
the insulator with increasing $U/W$ \cite{Brinkman}.

\begin{figure}
\includegraphics[clip,width=3.25in]{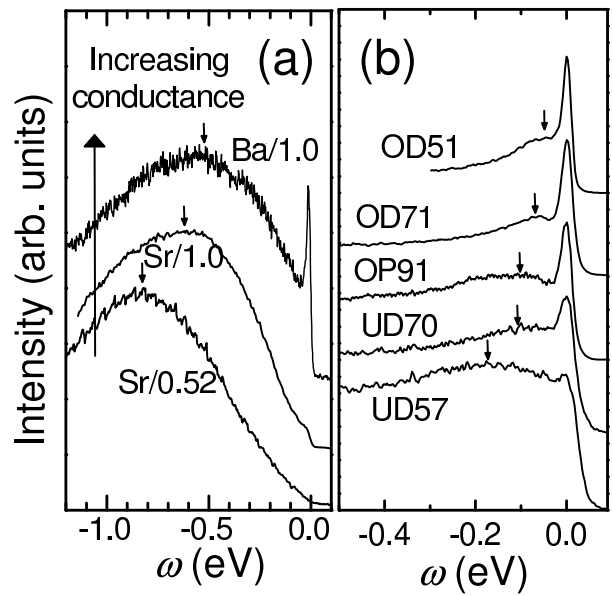}
\caption{(a) Broad EDC spectra of cobaltates, samples and temperatures as in
Fig. 1. (b) EDC spectra of Bi2212 cuprates near the ($\pi,0$) region in the
superconducting state. UD69 is underdoped cuprate with $T_{c}=69$K, OP91 is
optimally-doped with $T_{c}=91$K, and OD51 is overdoped with $T_{c}=51$K.
For comparision, each spectra have been normalized to the energy and intensity
of the sharp peak. In both panels, the arrows point to the visual estimate of
the location of the maximum broad peak.}
\label{fig2}
\end{figure}

Fig. 2a displays the EDC's of the same samples from Fig. 1 over a
broader energy range and in a manner suitable for sample to sample
comparison. The position and extent of the broad, incoherent part
of the excitation is more easily seen here. The spectra are
arranged in order of increasing room temperature conductance with
the bottom spectra ($Sr/0.52$) the least conducting, $Sr/1.0$ in
the middle, and $Ba/1.0$ on top the most conducting. From the
bottom spectrum to the middle, the predominant change is an
increase in the number of carriers. From the middle spectrum to
the top, the predominant change is a decrease in $U/W$. The broad
peak moves toward $E_{F}$ with increasing conductivity regardless
of the nature of the doping involved. The largest change is
associated with adding charge carriers, presumably connected with
a change in the chemical potential as in semiconductors. As
mentioned above, changing the number of carriers in samples that
are already basically conducting, there is a small change in the
position of the incoherent states but a large increase in the
spectral weight of the sharp, quasiparticle related peak at
$E_{F}$.

There is a close connection between the spectral lineshape at
$E_{F}$ and the resistivity measurements. For $Sr/0.52$, the
spectra consist of the broad, incoherent peak and a lack of any
intensity at $E_{F}$. This corresponds with the insulating
behavior in all directions of the temperature-dependent
resistivity \cite{Loureiro,Tsukuda}. In $Sr/1.0$, the spectra is
still dominated by the broad incoherent peak, but a finite
intensity develops at $E_{F}$ with a weak coherent peak. As shown
in Fig. 3(b) this sharp peak diminishes with increasing
temperature, and is no longer detected at 130K. Fig. 3(a) shows
that in-plane $\rho_{ab}$ is metallic over the temperature range
studied while $\rho_{c}$ shows predominantly insulating behavior
with a crossover to weakly metallic behavior below 120K. The
downturn in $\rho_{c}$ corresponds to the crossover temperature
where the sharp peak in the ARPES spectra appears. Fig. 3(c) and
(d) show similar behaviour, but with more pronounced peaks and a
more pronounced turnover to metallic temperature dependence for
$\rho_{c}$. These characteristics are consistent with that
described in our earlier paper \cite{Valla-Nature} which included
measurements on a $Ba/1.0$ type sample. In both metallic samples,
the presence of the coherent, sharp peak in the spectra
corresponds with the metallic behavior of $\rho_{c}$, whereas the
metallic behavior of $\rho_{ab}$ corresponds to the presence of a
non-zero intensity in the spectra at the $E_{F}$.

Our previous paper established a connection between the
temperature-dependent behavior of $\rho_{ab}$ and $\rho_{c}$ with
the effective dimensionality of the system \cite{Valla-Nature}. A
3D system is defined as having $\rho_{ab}$ with similar
temperature dependence as $\rho_{c}$,i.e. $\rho_{ab}/\rho_{c}$ is
constant. Hence, sample $Ba/1.0$  is essentially 3D below 150K
since both $\rho_{ab}$ and $\rho_{c}$ have similar behavior
($d\rho/dT > 0$). On the other hand, above 200K, the sample shows
2D properties since $\rho_{ab}$ and $\rho_{c}$ have opposite
behavior. The nature of the charge carriers in this 2D phase is
unknown but appears to not be independent quasiparticles. In all
cases the lowest temperature phase appears to be 3D in nature,
whether metallic or insulating. This is reminiscent the of
high-$T_{c}$ superconductors such as La$_{2-x}$Sr$_{x}$CuO$_{4}$
where measurement of the resistivity at low temperatures in the
normal state induced by large magnetic fields, has shown
$\rho_{ab}$ and $\rho_{c}$ to turn insulating for less than
optimal doping \cite{Boebinger}. Further, in those samples, for
any doping $\rho_{ab}$ and $\rho_{c}$ are either both insulating
or both metallic at lowest temperature.

\begin{figure}
\includegraphics[clip,width=3.25in]{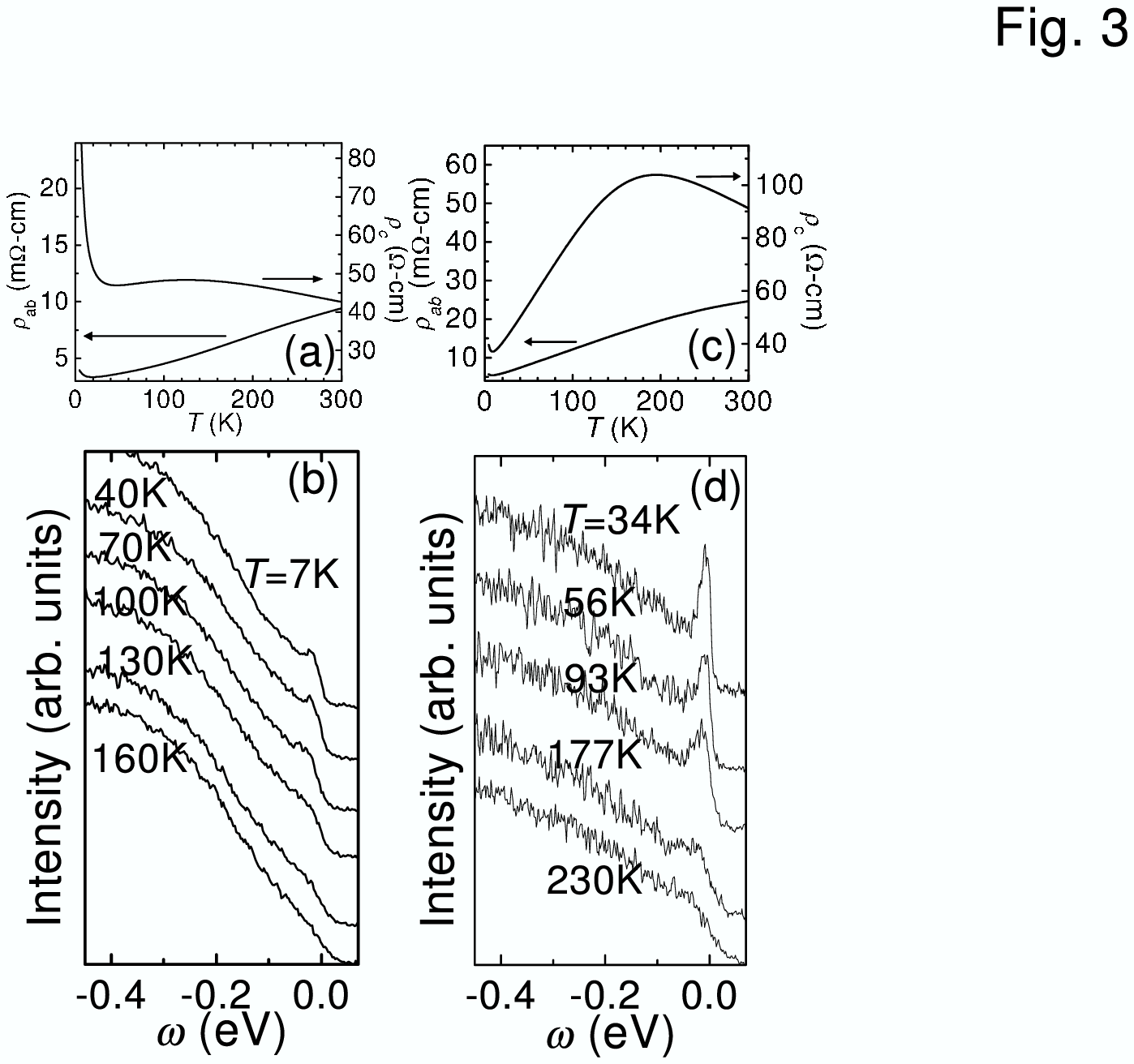}
\caption{Temperature dependent behavior of $\rho_{ab}$ and
$\rho_{c}$ (top panels), and the ARPES spectra (bottom panels) for
Samples $Sr/1.0$ (Fig (a) and (b)) and $Ba/1.0$ (Fig. (c) and
(d)).}
\label{fig3}
\end{figure}

We are not aware of a model that explicitly predicts a two-part
MIT as a function of temperature. However, the ARPES spectra
themselves have a strong resemblance to the spectral functions
calculated within the frame work of the infinite-dimensional DMFT
for a reasonable range of parameters \cite{Rozenberg,Pruschke}. A
comparison of Figs. 7-10 vs. 12-15 of Pruschke et al. \cite{Pruschke} show
the MIT as a function of carrier concentration. Pruschke et al.
give the spectral functions in fig. 7 (for half filling and U=3)
and Fig. 12 (for hole doping of 0.03 and U=3). The experimentally
similar variation is going from the spectra of $Sr/0.52$ to
$Sr/1.0$. The differences are that our lower level of doping is
not half filling and U in our samples is enough larger that there
is a gap in the lesser-doped material. Aspects of our data which
seem to match the calculation include the ubiquitous presence of a
broad high energy excitation, a tendency for the broad peak to
sharpen and move to lower energy with increasing conductivity, and
the formation of a slowly dispersing sharp peak in the metallic
state. The agreement for $Sr/1.0$ extends to the region of low
filling and high temperatures in which the calculated spectra have
states at $E_{F}$ but no QP peak. Further, changes between
$Sr/1.0$ and $Ba/1.0$ are largely a function of changing $U/W$. 
While it is difficult to find a study that explicity examines the spectral
function as a function of $U/W$ away from half filling, the
trends can be seen in studies at half filling. Both Fig. 16 and
Fig. 18 of reference \cite{Georges} show the evolution of the
spectral function as a function of $U/W$. The trends match our
data in showing a substantial increase in the spectral weight of
the sharp peak near $E_{F}$ with decreasing $U/W$ as well as
significant lowering of the energy position of the broad,
incoherent peak. The most striking agreement between our data and
the DMFT calculated spectral functions is for the temperature
dependence of our most conducting sample, $Sr/0.52$ as shown in
Fig 3d here and Fig. 2 of \cite{Valla-Nature}, compared to Fig. 12
of Ref. \cite{Pruschke}.

The connection between the conductivity data and the DMFT
calculations is less clear. The calculations indicate that the
disappearance of the sharp peak in the density of states with
increasing temperature corresponds to a transition to insulating
isotropic resistivity at half filling (Figs. 12 and 14 of Ref.
\cite{Pruschke}). This what we observe for $\rho_{c}$ for the
metallic phases \cite{Valla-Nature}. The measured behavior of
$\rho_{ab}$, i.e. that it remains metallic with no indication of
the crossover, is not present in the DMFT calculations.

There are similarities between the spectral function for the
cuprate superconductor Bi2212 and these cobaltates. Unlike the
cobaltates, the cuprate spectral lineshapes are strongly
anisotropic within the plane, at least for optimal or lesser
doping. The cuprate spectra in the vicinity of the M-point are
qualitatively similar to the cobaltates \cite{cuprate}. In Fig.
2b, spectra from Bi2212 at different doping levels: underdoped
($T_{c}=69K$), optimally doped ($T_{c}=91K$) \cite{Fedorov}, and
overdoped  ($T_{c}=51$K) \cite{Yusof}. The cuprate spectra are
taken in the superconducting state near the M-point of the
Brillouin zone. The spectra are arranged such that the top spectra
comes from the most conducting compound (OD51) while the bottom
spectra is from the least conducting (UD69) \cite{Kendziora}. The
two overall patterns that were observed in the cobaltates are
repeated here for the cuprates. With increasing conductivity of
the sample, the broad hump sharpens and shifts to lower energies,
while the intensity of the sharp peak increases. In both cases the
spectral evolution is consistent with that expected from DMFT
calculations and it appears that the appearance of QP states
corresponds with the emergence of isotropic electronic behavior.

A possible caveat to the above observation is that some literature
has attributed the broad peak near the M-point in Bi2212 as being
due to a second band of the bilayer CuO planes \cite{bilayer}.
While there are few band calculations that explicitly consider
charge doping, indications are that the splitting of the bilayer
bands should be roughly independent of doping \cite{Lichtenstein}.
Since ARPES results show that the energy difference between the
sharp peak and the broad peak clearly shrinks with doping, and the
relative spectral weight of the broad peak continuously decreases,
we conclude that this broad peak is not predominantly a second
band. On the other hand, the evolution of this broad low energy
peak is consistently predicted if this feature is predominantly
the incoherent part of the excitation. In addition, at least one
calculation that considers the effects of bilayer split bands
explicitly predicts that this broad feature is predominantly due
to the incoherent excitation \cite{Eschrig}.

The general picture put forward here concerning the relationship
between the sharp quasiparticle peak and the broad incoherent peak
it emerges from is extremely similar in spirit to a recent work
concerning the evolution of the cuprate compounds from Mott
insulator to superconductor \cite{KShen}. However, that work does
not address the temperature evolution. Another system for which
comparisons are naturally drawn are the compounds of the type
$Na_xCoO_2$. For doping near x=0.35 these are the precursors to
the water intercalated superconductors with $T_C = 5 K$. Those
compounds consist of edge sharing $CoO_6$ octahedra forming
electronically active triangular sheets separated by Na layers and
water in the hydrated versions. Thus those compounds have similar
structure to the cobaltate compounds discussed here in many
aspects. For heavily doped samples $Na_{0.7}CoO_3$ the electronic
structure near the Fermi level is very similar to that reported
here for the most conducting compounds, $Ba/1.0$ \cite{Hasan1}. At
low temperatures there is a predominantly circular Fermi surface,
a broad peak near 0.7 eV about $0.6 eV$ in width out of which a
small, sharp peak emerges that is weakly dispersive. This was
labelled as the quasiparticle peak similar to our own
identification. The temperature dependence of that quasiparticle
peak is also similar to what we detect, emerging below $T = 120
K$. In that case it was reported that this temperature coincides
to a change in the in-plane resistivity from T-linear at low
temperature to a stronger dependence at high temperature. The out
of plane resistivity was not reported. Interestingly, the lower
doped $Na_{0.3}CoO_3$ the equivalent quasiparticle peak has a much
greater dispersion, perhaps indicating that $U/W$ is much smaller
for this latter compound \cite{Hasan2}. Another group reporting on
$Na_{0.6}CoO_3$ found dispersion of the quasiparticle peak that
was fairly broad, closer to that of $Na_{0.3}CoO_3$ \cite{Ding}.
The exact behavior of this peak as a function of doping should
prove very interesting.

In summary, we have shown ARPES results for the bismuth cobaltates
across the MIT. The correlated insulators have broad,
non-dispersing, incoherent spectra. The nominally metallic samples
are more complex with at least two phases; (i) metallic behavior
in the plane accompanied by the appearance of a Fermi edge in an
otherwise broad spectral function and (ii) at lower temperatures a
3D electronic state occurs accompanied by the transfer of spectral
weight to a sharp, slowly dispersing quasiparticle peak. As
conductivity increases from sample to sample, either through
increasing the number of holes or decreasing $U/W$, the broad
incoherent part of the spectrum sharpens and shifts towards the
Fermi energy and the sharp peak increases in intensity. The
spectral function near the $\Gamma$M region of the Bi2212
superconducting compounds evolves similarly. Despite the
prominence of a phase that is metallic in-plane and insulating
out-of-plane, the zero temperature phase for all of the layered
compounds appear to be 3D in nature.

We acknowledge valuable discussions with V. Perebeinos, A.
Tsvelik, J. Tu, and R. Werner. Work supported in part by Dept. of
Energy under contract number DE-AC02-98CH10886, DE-FG02-00ER45801,
and DOE-BES W-31-109-ENG-38, and BOW acknowledges the support of a
Cottrell Scholar Fellowship.


\begin{references}
\bibitem{cuprate} We note that the cuprate electronic structure is highly
anisotropic. Only the cuprates spectra along the Cu-O bond
direction ($\Gamma$M) show similar lineshapes with the cobaltates.
This is the region that determines the $c$-axis transport of the
cuprates.
\end{references}
\end{document}